# A ferro-deformation at the open quantum system with protons, Z = 8, neutrons, N = 20: $^{28}$O

Chang-Bum Moon*

*Hoseo University, Chung-Nam 336-795, Korea*

We offer a possibility that the nuclear system with protons, Z = 8 has a large deformation at neutrons, N = 20; $^{28}$O that is beyond the neutron drip line. According to our previous works [arXiv: 1604.05013, 1604.02786, 1604.01017], it is expected that the ferro-deformation would occur at Z = 8, N = 20 through a shape phase transition at N = 18 out of N =16. The shape transition can be explained in terms of isospin dependent spin-orbital interactions between neutrons in the $d_{3/2}$ orbital and protons in the $d_{5/2}$ orbital, by yielding both the neutron and the proton pseudo-shell configurations built on each combined subshells. We argue that such a large deformation at N = 18 would be responsible for the $^{26}$O to be unbound, leading to a ground state neutron emitter. The ferro-deformation is mapped on the nuclear chart such that is around the following critical proton, neutron coordinates, (Z, N); (64, 104), (40, 64), (20, 40), (8, 20). This configuration depicts a beautiful pattern coming from a harmonious order in the microscopic quantum world.

**Keywords**: national nuclear data center (NNDC), nuclear shell closure, magic numbers, pseudo-shell, shape phase transition, ferro-deformation, isospin dependent spin-orbital interaction.

*cbmoon@hoseo.edu

## 1. Introduction

In the history of quantum mechanics, the nuclear magic number is famous for indicating the existence of shell structure. This magic number corresponds to the nucleon number; protons, Z, and neutrons, N, at a large shell closure like the periodic numbers of the inert gases group in the periodic table of the elements. The magic numbers in the nuclei could be reproduced by introducing a strong spin-orbital coupling term into the single particle potential [1, 2]. It implies that the spin-orbital interaction in a nucleus plays a decisive role in controlling nuclear structures. Figure 1 illustrates the nuclear shell model scheme that produces the magic numbers, such as 2, 8, 20, 28, and 50. In the practical experimental observables, based on the first $2^+$ excited energies, E($2^+$), in Fig. 1(b), we can see such a large energy gap at N = 8, 20 and 28. It should be noticed that the numbers 8, 20, and 40 are also shell gap numbers built on the harmonic oscillator potential, as shown in Fig. 1(a) [3].

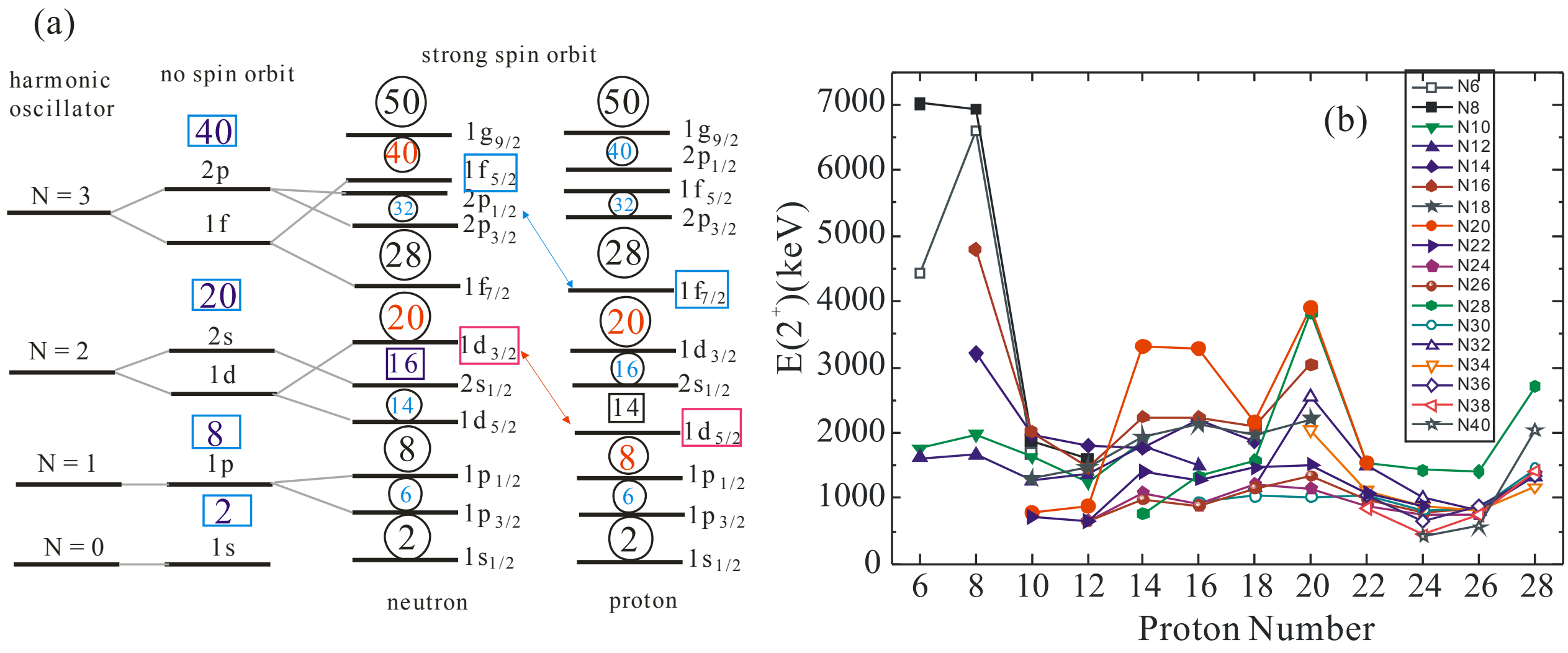

Fig. 1. (a) The single particle energies of a harmonic oscillator potential as a function of the oscillator quantum number N, the single-particle energies of a Woods-Saxon potential, and the level splitting due to the spin-orbit coupling term. Note that the levels are numbered serially by a given orbital quantum number according to Nj = 2j + 1 of nucleons that can occupy each state. For discussions, we denote the associated spin-orbit doublet by the arrow and the corresponding critical points of Z = 8 (20) and N = 20 (40) are in red(blue) color. Also notice those reinforced shell gaps at Z = 14 and N = 16. (b) Systematics for the first $2^+$ excited states in the nuclides over Z = 6 to 28 as a function of neutron numbers. Data are taken from NNDC [4].

In nuclear physics, the nucleus $^{16}$O, with Z, N = 8, is known as a representative double-magic nucleus revealing the highest $2^+$ excited state all over the chart of the nuclides. Over last two decades, searching for the existence of the double-shell gap at Z = 8, N = 20 were being considerably made. However, this system, including the Z = 8, N = 18 system has turned out to be neutrons radioactivity from the ground state, indicating the Z = 8, N = 16 is the last configuration in oxygen nuclides before the neutron drip line [4-11]. This work aims at describing nuclear structures of the $^{28}$O, with a motivation by addressing a question whether it has *a double-shell gap or deformation* in the ground state. Following systematic studies as in the previous works [12-14] and in the framework of the isospin-dependent spin-orbital interactions, we suggest that at Z = 8, N = 18 occurs a phase transition, yielding a large deformation from the spherical system $^{24}$O. In turn, the system with Z = 8, N = 20 has a ferro-deformation, as in cases of the Z = 20, N = 40 [14], Z = 40, N = 64 [13], and Z = 64, N = 104 [12]. As a conclusion, we argue '*a reason why the Z = 8 system is unbound at N = 18'* would be related with such a sudden deformation.

## 2. A ferro-deformation at Z = 8, N = 20.

For the present work, we start to discuss the systematic behavior of low-lying level properties based on the excitation energy ratio of the first $2^+$ and $4^+$ states, R = $E(4^+)/E(2^+)$. in the even-even nuclei. This value indicates a deformation parameter such that R < 2 for a spherical nucleus, R ~ 2 for a vibrator and R ~ 3.3 for a deformed axial rotor [15, 16]. Figure 2 illustrates the systematics of the R values, including the $E(2^+)$, within Z = 8 and 28, along N = 8 to 44.

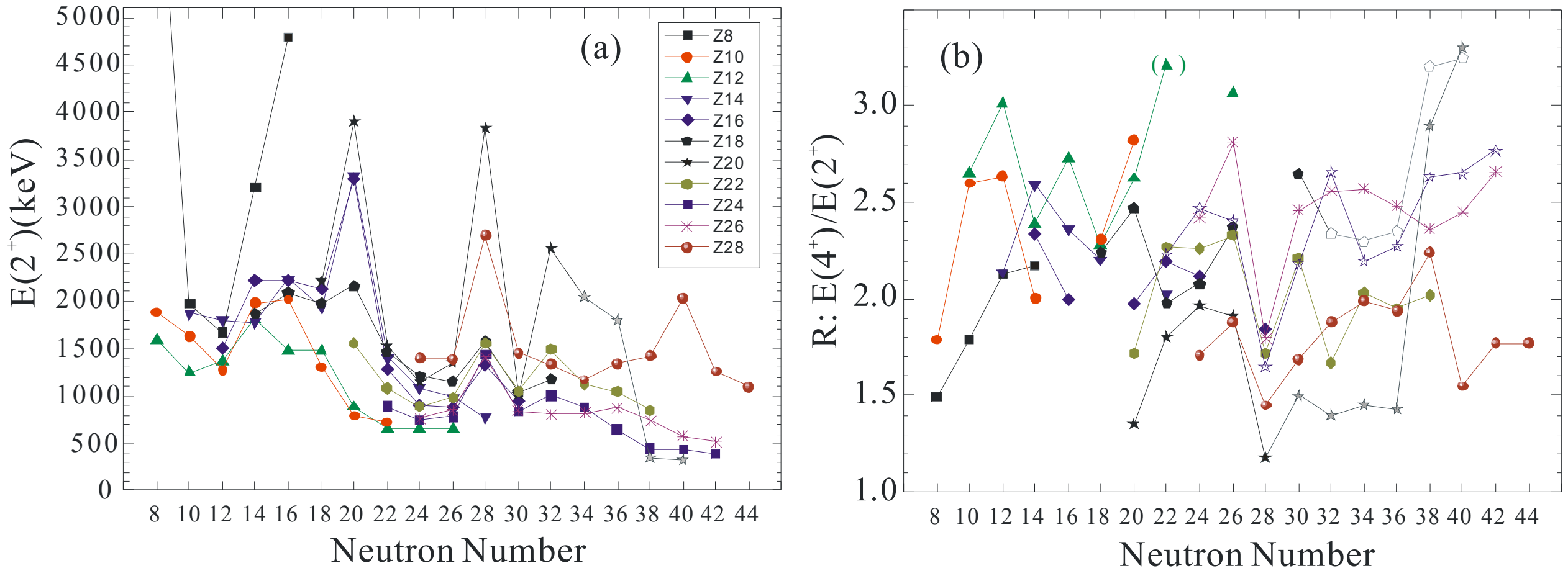

Fig. 2. (a) Systematic plots of $E(2^+)$ and (b) R, $E(4^+)/E(2^+)$, values as a function of neutron numbers for the nuclei within the Z = 8 and 28 space. Data are from NNDC [4]. The point at Z = 12, N = 22 is not certain at the current NNDC.

In both R and $E(2^+)$ systematics, we notice firstly a large shell gap at N = 14, 16, 32, 40, in addition to traditional shell gap numbers, N = 8, 20, 28. Second is that, though N = 20 is a magic number, such a large shell gap disappears at N = 20 with Z = 10; $^{30}$Ne, indicating a good rotator with R ~ 2.8. This feature provides us a hint that the Z = 8 has a possibility for having a deformation at N = 20. For the case of protons Z = 20, we know that at N = 40 there appears a large(ferro) deformation [14]. We can see such a ferro-deformation in Fig. 2(b). From comparisons of both parameters in the two regimes, one is Z = 8, N = 20, and the other Z = 20, N = 40, we could deduce a critical point that would take place a shape phase transition in the Z = 8, N = 20 region. With concepts of the pseudo-shell configuration, built on the combined subshells as shown in the previous works [12, 13, 14], we are able to expect a critical point at which the ferro-deformation occurs.

Table 1. Summary of pseudo-shell configurations between double-shell closures of 2 ≤ Z ≤ 14 and 8 ≤ N ≤ 28: pseudo-shells, combined subshells, total spins with the capacitance of nucleon numbers, contributing nucleon numbers to deformation, and corresponding deformation configurations. It is worthwhile to know that the nucleon numbers contributing to a ferro-deformation is related with the Hund rule.

| pseudo-shells | combined subshells | pseudo-shell total spins: nucleon capacitance | nucleon number contributing to ferro-deformation | Corresponding summed nucleon number | representative deformation configurations: (Z, N) |
|---|---|---|---|---|---|
| **$J_{pd}$(11/2)** | $1p_{3/2}1p_{1/2}1d_{5/2}$ | 11/2: 12 | 6 | 8 | (8, 20), (8, 18) |
| $J_{pd}$(11/2) | $1p_{3/2}1p_{1/2}1d_{5/2}$ | | 4 | 10 | (10, 20), (10, 22) |
| **$J_{sdf}$(13/2)** | $2s_{1/2}1d_{3/2}1f_{7/2}$ | 13/2: 14 | 6 | 20 or 22 | (8, 20), (10, 22) |
| $J_{sdf}$(13/2) | $2s_{1/2}1d_{3/2}1f_{7/2}$ | | 4 | 18 | (8, 18) |

In the proton region, 2 $\leq$ Z $\leq$ 20, the pseudo-shell, $J_{pd}(11/2)$ built on the combined subshells $1p_{3/2}1p_{1/2}1d_{5/2}$ has the capacitance of nucleons 12 while in the neutron region, 8 $\leq$ N $\leq$ 28, the $J_{sdf}(13/2)$ built on the $2s_{1/2}1d_{3/2}1f_{7/2}$ has the 14 capacitance. We know that a reinforced deformation arises when both proton and neutron pseudo-shells are closely half-filled. As with the previous studies for the ferra-deformation at the critical point Z = 20, N = 40 [14], it is expected that a shape phase transition should occur at Z = 8 when N = 16 to 18. In turn, a ferro-deformation is formed at the critical point, Z = 8, N = 20. Following this scenario and the systematics based on the $E(2^+)$ and R parameters in the region of Z = 20, N = 40, we can assign the $E(2^+)$ and R values into the Z = 8 with N = 18 and 20.

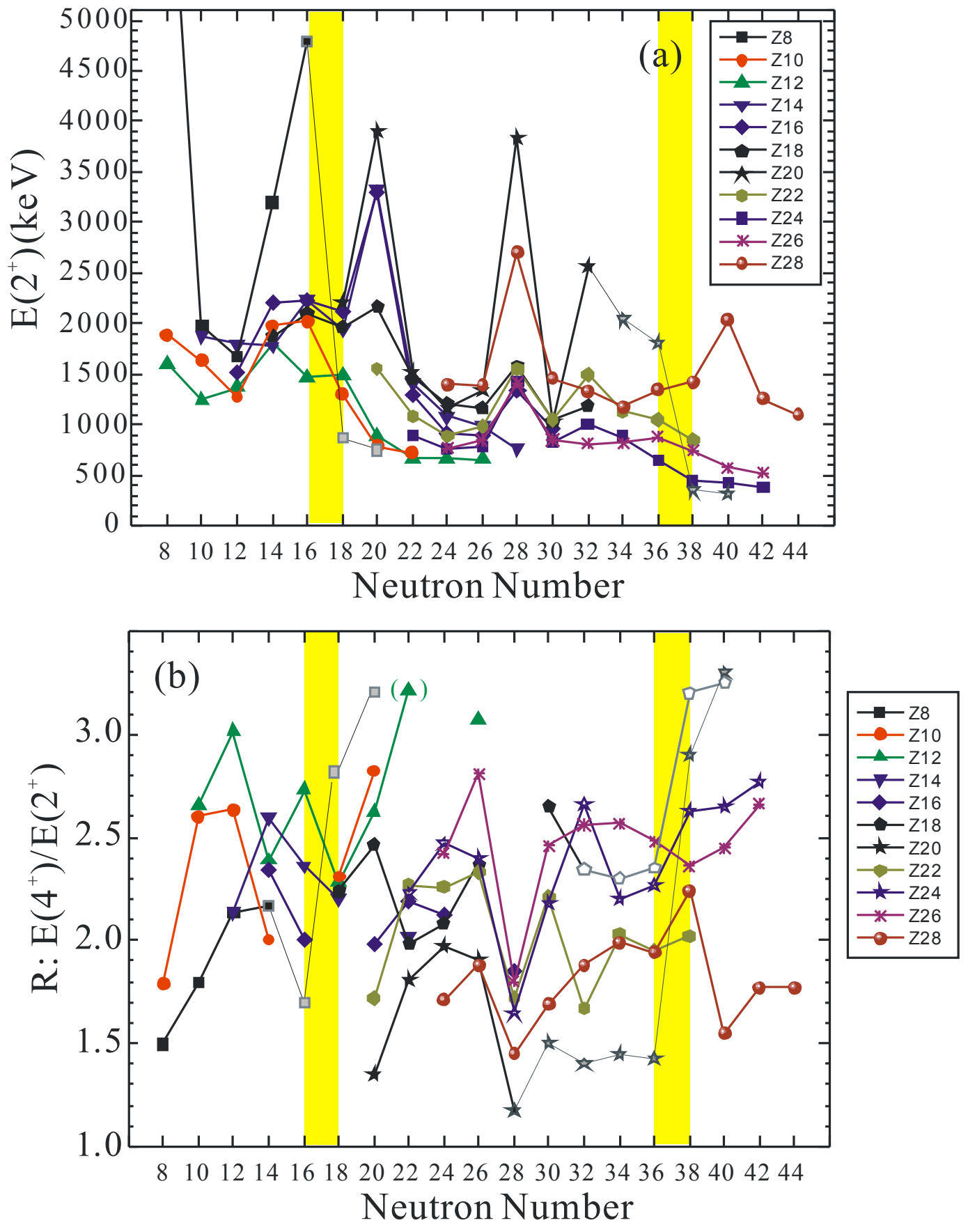

Fig. 3. (a) Systematic plots of $E(2^+)$ and (b) R, $E(4^+)/E(2^+)$, values as a function of neutron numbers for the nuclei within the Z = 8 and 28 space. The shaded regions include the critical points that shape phase transitions take place. The points connected with dotted lines correspond to the expected values.

For further confirming the pseudo-shell concepts, we introduce a (Z, N) configuration contributing to a ferro-deformation, indicating what amounts of nucleons are occupied one by one in each pseudo-subshell, $J_{pd}(11/2)$, $J_{sdf}(13/2)$. For instance, the $J_{pd}(11/2)$ pseudo-shell is half-filled at protons 6 while the $J_{sdf}(13/2)$ at neutrons 7 ( 6 in the case of even number). Accordingly, the configuration (Z = 6, N = 6) forces the most strong deformation. Here we define it a Hund configuration, by recalling the so called Hund's rule in atomic physics [17]. It is instructive to point out that the Hund configuration, (Z = 6, N = 6) corresponds to the system Z = 8, N = 20; $^{28}$O. The next configuration is (6, 4) or (4, 6), which does to the Z = 8, N = 18, $^{26}$O, or to the Z = 10, N = 20, or 22; $^{30}$Ne, $^{32}$Ne. Third is (4, 4), which does to the Z = 10, N = 18, $^{28}$Ne, or to the Z = 10, N = 24, $^{34}$Ne. We find that the $^{30}$Ne, Z = 10, N = 20, with the Hund configuration (4, 6), has R ~ 2.82 and the $^{28}$Ne, Z = 10, N = 18, with (4, 4) has R ~ 2.3. Following this systematic, we assign the Z = 8, N = 18, with (6, 4), to be R = 2.82. As shown in Table 1, we summarize the Hund configuration as the nucleon numbers contributing to a deformation such that; Z = 8, N = 20: (6, 6), Z = 8, N = 18: (6, 4), Z = 10, N = 20 : (4, 6), Z = 10, N = 22: (4, 6), Z = 10, N = 18: (4, 4). We again emphasize that they are parameters indicating how many nucleons are aligned one by one in each pseudo-subshell for yielding a deformation.

We also expect that the first $2^+$ energy at Z = 8, N = 18 would be about 900 keV. This result is obtained from a comparison with the $2^+$ energy at Z = 20, N = 36 and that at Z = 20, N = 38 [14], which marks 5 times. In other words, the energy 900 keV is about fifth times the $2^+$ energy at Z = 8, N = 16, $^{24}$O. Next, the $4^+$ excitation energy for the Z = 8, N = 18 comes from the R value 2.82. For the case of Z = 8, N = 20, taking the Hund (6, 6) configuration into account, we assign a value of R to be 3.2, which is slightly lower than that at Z = 20, N = 40 [14]. On the other hand, we notice that the R value at Z = 12, N = 22, $^{34}$Ge, indicates a large deformation, being likely a ferro-deformation. We argue that the deformation at the Z = 12 system is different in mechanism from the ferro-deformation. We will discuss this issue in the forthcoming paper [18]. Our results are summarized in Figs. 3 and 4 where the predicted values for the Z = 8 system are drawn with the dotted lines. In addition, we provide our proposed level schemes for the $^{26}$O and $^{28}$O, including the $^{24}$O in Fig. 5.

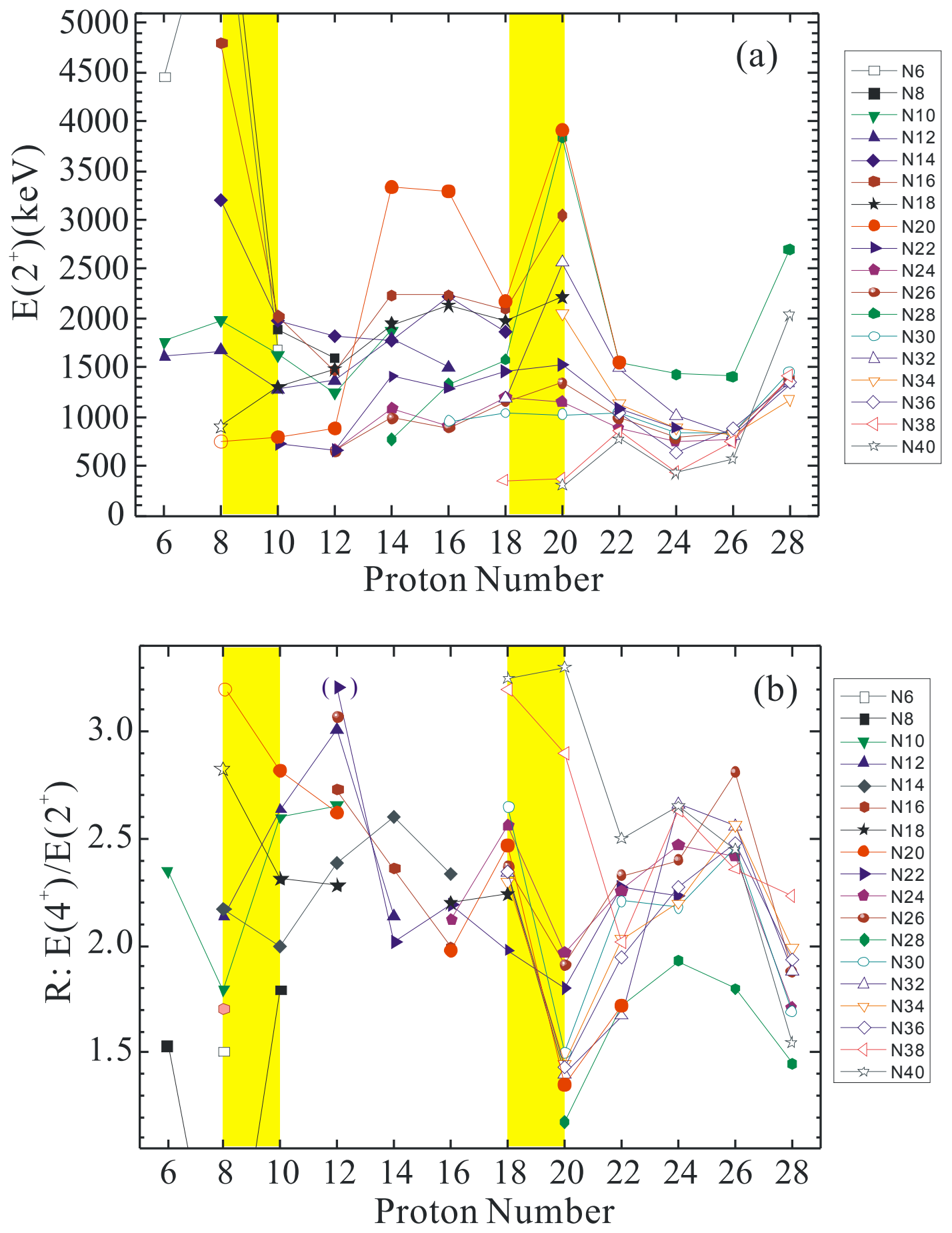

Fig. 4. Plots for the systematics of E($2^+$) and the R (= E($4^+$)/E($2^+$)) values between the N = 6 to 40 space as a function of proton numbers. The shaded regions include the critical points that shape phase transitions take place. The points connected with dotted lines correspond to the expected values.

The ferro-deformation can be explained within the framework of the spin-orbital interactions between protons and neutrons in the spin-orbital doublet; the $d_{3//2}$ for neutrons, and the $d_{5/2}$ for protons. See Fig. 1(a). A sudden shell structure change when N = 16 to 18 is caused by this isospin dependent spin-orbital interaction, which induce a huge neutron pseudo-shell, $J_{sdf}(13/2)$, along with a proton pseudo-shell $J_{pd}(11/2)$. Eventually, under the Hund (6, 6) condition, a maximum deformation comes with the ferro-deformation at the critical point; Z = 8, N = 20.

In Fig. 1(a), one can see more widen shell gaps, owing to the neutron $d_{3/2}$-proton $d_{5/2}$ doublet interaction, at N = 16 and Z = 14. Interestingly, the semi-shell gap at N = 16 is reinforced with Z = 8 while that at Z = 14 reinforced with N = 20.

They say that $^{24}$O has a double shell gap with *a new magic number* N = 16. Following this, we have to say that $^{34}$Si has a double-shell gap with *a new magic number* Z = 14. Notice the very dramatic shape phase transitions within the space $8 \leq Z \leq 14$ and $16 \leq N \leq 20$, which is governed by the neutron $d_{3/2}$ and proton $d_{5/2}$ spin-orbital doublet interaction. It is necessary to approach the nuclear quantum world with *a universal and holistic point of view*.

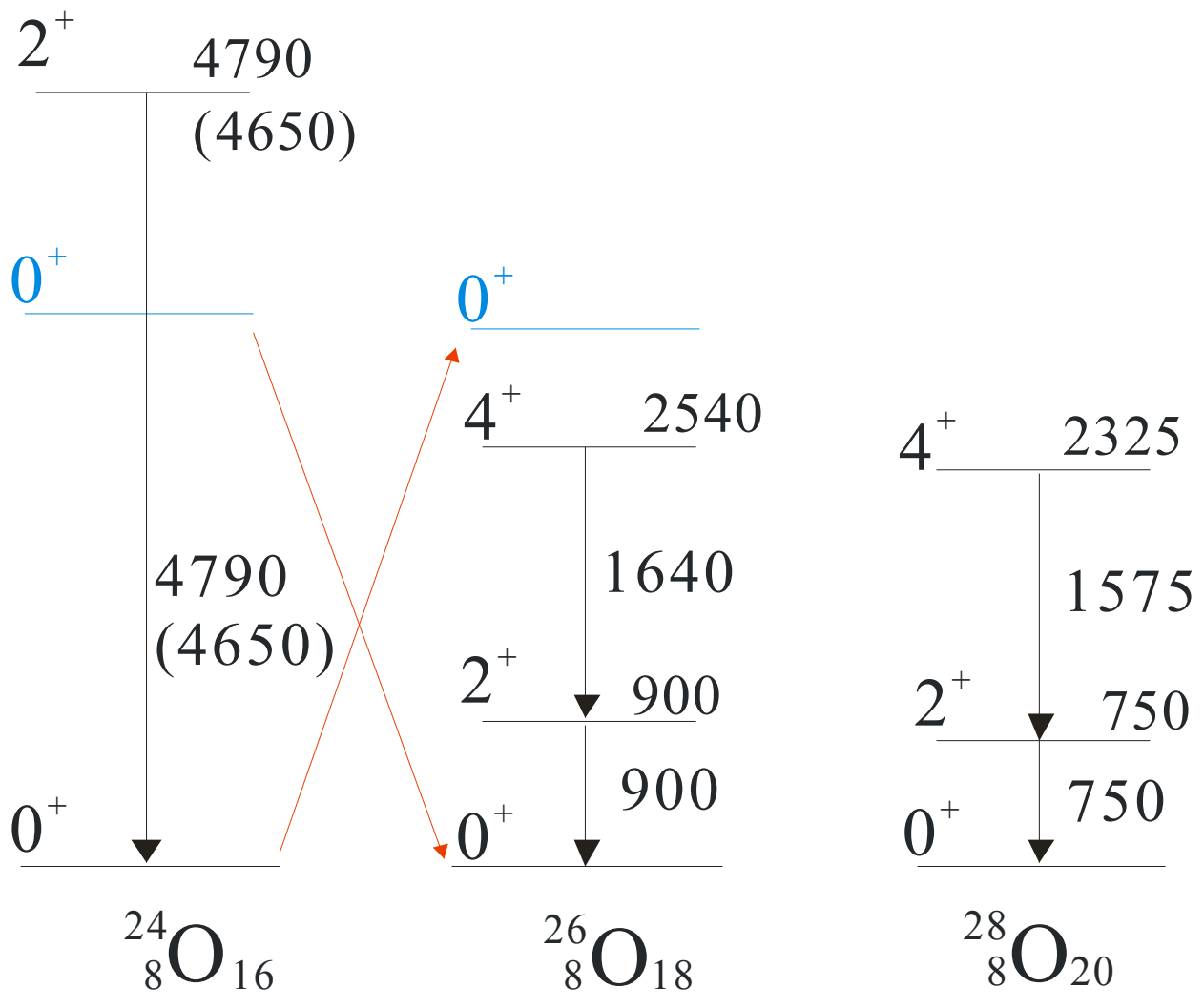

Fig. 5. The proposed level schemes for the Z = 8, with N = 16, 18, and 20. It is noted that an assumed second $0^+$ state is added for indicating possible shape coexistence. We suppose that the uncertainty for the $2^+$ transition energy to the ground state would be within 20 % in each nucleus. The $2^+$ energy in parenthesis of $^{24}$O is taken from [11]. Notice that by taking R ~ 1.7 for $^{24}$O, as shown in Figs. 3(b) and 4(b), the $4^+$ excitation energy would be around 8 MeV.

Before closing this section, we would like to remind of a big question to be still answered, '*why should the* $^{26}$O *be a ground state neutron emitter* ?' An answer can be made by the fact that the Z = 8 system should be deformed *in the ground state* through a sudden phase transition at N = 18 out of N = 16. Such a dramatic change in shape would prohibit forming the bound $^{24}$O + 2n system.

**4. Conclusions**

We suggested that the neutrons-unbound system with Z = 8, N = 20, $^{28}$O, has a large deformation. As in our previous works [12, 13, 14], we were able to demonstrate the occurrence of a reinforced deformation at Z = 8, N = 18 through a shape phase transition. The spin-orbital interaction between neutrons in the $d_{3/2}$ orbital and protons in the $d_{5/2}$ orbital induces such a reinforced deformation and finally yields the ferro-deformation at Z = 8, N = 20. In this circumstance, all the subshells in the region of $2 \leq Z \leq 14$ for protons and of $14 \leq N \leq 28$ for neutrons are combined to build the pseudo-shell such that $J_{pd}(11/2)$ from the $1p_{3/2}1p_{1/2}1d_{5/2}$ and $J_{sdf}(13/2)$ from the $2s_{1/2}1d_{3/2}1f_{7/2}$, respectively The ferro-deformation could be complete under both half-filled conditions in such a way that the Hund rule keeps. We argue that such a large deformation at N = 18 would be responsible for the $^{26}$O to be a ground state neutron emitter. The ferro-deformation is mapped on the nuclear chart over the following critical proton, neutron coordinates, (Z, N); (64, 104), (40, 64), (20, 40), (8, 20).

We expect that shape coexistence should be at Z = 8, N = 16 and Z = 8, N = 18, including the ferro-deformation at a higher lying level built on the second $0^+$ state in $^{24}$O. We hope that searching for the $2^+$, $4^+$ and the second $0^+$ states in $^{24}$O, $^{26}$O, and $^{28}$O would be made at the current rare isotopes beam facility; the Facility for Rare Isotope Beams (FRIB) at MSU in the U.S.A, the Radioactive Isotope Beams Factory (RIBF) at RIKEN in Japan, the Fragment Separator (FRS) at GSI in Germany, and HIE-ISOLDE at CERN in Europe.

**Note added 1**: The present results have been made with phenomenological arguments based on the systematics of experimental observables given by the national nuclear data center, NNDC. Accordingly, many references, possibly related to the present work in the literature could not be quoted because of the referred data from NNDC.
**Note added 2**: Any abbreviation is avoided since it makes the readers confusing like a jargon.
**Note added 3**: During this work, we come to know that a paper reporting the $2^+$ excited state of $^{26}$O has been published [19].

**References**

[1] M. G. Mayer, Phys. Rev. 75, 1969 (1949).
[2] O. Haxel, J. H. D. Jensen, and H. E. Suess, Phys. Rev. 75, 1766 (1949).
[3] Chang-Bum Moon, AIP Adv. **4**, 041001 (2014).
[4] National Nuclear Data Center, Brookhaven National Laboratory, http:// www.nndc.bnl.gov/ (February 2016).
[5] D. Guillemaud-Mueller, J. C. Jacmart, E. Kashy, A. Latimier, A. C. Mueller, F. Pougheon, and A. Richard, Yu. E. Penionzhkevich, A. G. Artuhk, A. V. Belozyorov, and S. M. Lukyanov, R. Anne, P. Bricault, C. Détraz, M. Lewitowicz, and Y. Zhang, Yu. S. Lyutostansky and M. V. Zverev, D. Bazin, W. D. Schmidt-Ott , Phys. Rev. C **41**, 937 (1990).
[6] O. Tarasov, R. Allatt, J. C. Angklique, R. Anne, C. Borcea, Z. Dlouhy, C. Donzaud, S. Grevy, D. Guillemaud-Mueller, M. Lewitowicz, S. Lukyanov, A.C. Mueller, F. Nowacki, Yu. Oganessian, N.A. Orr, A.N. Ostrowski, R.D. Page, Yu. Penionzhkevich, F. Pougheon, A. Reed, M.G. Saint-Laurent, W. Schwab, E. Sokol, O. Sorlin, W. Trinder, J.S. Winfieldd , Phys. Lett. B **409**, 64 (1997).
[7] A. Schiller, T. Baumann , J. Dietrich, S. Kaiser, W. Peters and M. Thoennessen, Phys. Rev. C **72**, 037601 (2005).
[8] M. Thoennessen, Atomic Data and Nuclear Data Tables **98**, 43 (2012).
[9] Z. Kohley, A. Spyrou, E. Lunderberg, P. A. DeYoung, H. Attanayake, T. Baumann, D. Bazin, B. A. Brown, G. Christian, D. Divaratne, S. M. Grimes, A. Haagsma, J. E. Finck, N. Frank, B. Luther, S. Mosby, T. Nagi, G. F. Peaslee, W. A. Peters, A. Schiller, J. K. Smith, J. Snyder, M. J. Strongman, M. Thoennessen, and A. Volya, **arXiv**:1208.2969 (2012).
[10] C. R. Hoffman,, T. Baumann, D. Bazin, J. Brown, G. Christian, P. A. DeYoung, J. E. Finck, N. Frank, J. Hinnefeld, R. Howes, P. Mears, E. Mosby, S. Mosby, J. Reith, B. Rizzo, W. F. Rogers, G. Peaslee, W. A. Peters, A. Schiller, M. J. Scott, S. L. Tabor, M. Thoennessen, P. J. Voss, and T. Williams, Phys. Rev. Lett. **100**, 152502 (2008).
[11] K. Tshoo, Y. Satou, H. Bhang, S. Choi, T. Nakamura, Y. Kondo, S. Deguchi, Y. Kawada, N. Kobayashi, Y. Nakayama, K. N. Tanaka, N. Tanaka, N. Aoi, M. Ishihara, T. Motobayashi, H. Otsu, H. Sakurai, S. Takeuchi, Y. Togano, K. Yoneda, Z. H. Li, F. Delaunay, J. Gibelin, F. M. Marqu´es, N. A. Orr, T. Honda, M. Matsushita, T. Kobayashi, Y. Miyashita, T. Sumikama, K. Yoshinaga, S. Shimoura, D. Sohler, T. Zheng, and Z. X. Cao, Phys. Rev. Lett. **109**, 022501 (2012).
[12] Chang-Bum Moon, **arXiv**:1604.01017 (2016).
[13] Chang-Bum Moon, **arXiv**:1604.02786 (2016).
[14] Chang-Bum Moon, **arXiv**:1604.05013 (2016).
[15] K. S. Krane, Introductory Nuclear Physics (John Wiely & Sons, New York, 1988) pp. 136-137.
[16] K. Heyde, Basic Ideas and Concepts in Nuclear Physics (Institute of Physics Publishing, Bristol and Philadelphia, 1999) p. 361.
[17] N. W. Ashcroft and N. D. Mermin, Solid State Physics (Holt, Rinehart and Winston, New York, 1976) pp. 650-651.
[18] Chang-Bum Moon, to be appeared in **arXiv**:1605. (2016).
[19] Y. Kondo, T. Nakamura, R. Tanaka, R. Minakata, S. Ogoshi, N. A. Orr, N. L. Achouri, T. Aumann, H. Baba, F. Delaunay, P. Doornenbal, N. Fukuda, J. Gibelin, J.W. Hwang, N. Inabe, T. Isobe, D. Kameda, D. Kanno, S. Kim, N. Kobayashi, T. Kobayashi, T. Kubo, S. Leblond, J. Lee, F. M. Marqués, T. Motobayashi, D. Murai, T. Murakami, K. Muto, T. Nakashima, N. Nakatsuka, A. Navin, S. Nishi, H. Otsu, H. Sato, Y. Satou, Y. Shimizu, H. Suzuki, K. Takahashi, H. Takeda, S. Takeuchi, Y. Togano, A. G. Tuff, M. Vandebrouck, and K. Yoneda, Phys. Rev. Lett. **116**, 102503 (2016)